\documentclass[pra,twocolumn]{revtex4}
\usepackage{graphicx}
\usepackage{capt-of}
\usepackage{amsmath}

\begin{document}

%\draft

%\wideabs{

\title{Vortex-vortex interactions in toroidally trapped Bose-Einstein condensates}
\author{T. Schulte, L. Santos, A. Sanpera and M. Lewenstein}
\affiliation{Institut f\"{u}r Theoretische Physik, Universit\"{a}t Hannover,
30167 Hannover, Germany
}
\date{\today}

\begin{abstract}
We analyze the vortex dynamics and vortex--vortex interactions
in Bose-Einstein condensates confined in toroidal traps. 
We show that this particular geometry strongly distorts the vortex dynamics.
The numerically calculated vortex trajectories are well explained 
by an analytical calculation based on image method and conformal mapping. 
Finally, the dissipation effects are discussed.
\end{abstract}

\maketitle

\section*{Introduction}

% BEC and superfluidity.

The realization of Bose-Einstein Condensation (BEC) in trapped dilute alkali gases
 \cite{Rb87,Na,Li,H,Rb85,He} has opened the possibility to explore fundamental effects 
of low temperature physics in a very controllable way. 
In this sense, several effects related to superfluidity \cite{Landau} 
have been recently reported \cite{Oxford,MITsuper}.
Among the phenomena related to superfluidity, the possibility of observing  
vortices with quantized circulation \cite{Donnelly}
has aroused a big interest.  Due to the diluteness of the atomic condensates, 
accurate theoretical predictions are possible.
In addition, in these systems, the 
diameter of the vortex core is typically three orders of magnitude larger than for 
superfluid Helium, allowing for a more straightforward optical observation.
Numerous theoretical schemes have been suggested for the generation of vortices in BEC, 
see e.g. Refs.\cite{Martzlin97,Dum98,Jackson98,Caradoc99,Dobrek99,Williams99,Feder99}.
So far only two techniques have successfully generated vortices
in condensates, namely the stirring of a condensate 
with a blue detuned laser~\cite{Madison20,Abo2001}, and the phase imprinting method achieved via the coherent interconversion 
between the two components of a binary condensate~\cite{Matthews99}.
In addition the creation of a vortex ring after the decay of a dark soliton has 
been recently reported \cite{Anderson00}.
The dynamics and stability of vortices in trapped 
Bose--Einstein condensates has recently also been a subject of active investigation 
\cite{Fetterreview,shlyapnikov99,Guilleumas}. 

% Toroidal traps

In the majority of current BEC experiments, the condensate is created in a cylindrical trap, either pancake--shaped,  
or cigar--shaped. However, in the recent years the possibility of creating a BEC in toroidal 
traps has aroused a growing interest. Toroidal traps have already been used in early Sodium BEC experiments 
at MIT \cite{Na}, and in recent experiments on vortices at JILA \cite{Matthews99}. Also, very recently 
the idea of creating a toroidal optical dipole trap using Laguerre-Gaussian beams has been proposed  
\cite {Wright}. The physics of condensates in toroidal traps have been extensively investigated during the last recent 
years, including the ground state properties and elementary excitations \cite{Rokhsar,Salasnich}, the possibility 
to obtain vortices which due to the toroidal topology could be metastable 
beyond some critical nonlinearity \cite{Rokhsar,Salasnich,Benakli,Tempere},   
the generation and properties of solitons and the so called "svortices" in this kind of traps \cite{Brand}, and even the 
possibility to obtain a mode--locked atom laser with toroidal geometry \cite{Drummond}. 
Very recently, we have proposed that a time--averaged potential produced by the shaking of an usual Gaussian trap 
also leads to the formation of a double well or a toroidal condensate \cite{Martikainen}. Additionally, if during the condensate shaking 
a proper phase is imprinted, dark solitons and eventually vortices can be created in a controllable way within the 
torus. In Ref.\ \cite{Martikainen}, it was already reported, that the particular toroidal geometry significantly 
modifies the vortex dynamics, as well as the vortex--vortex interaction. It is the aim of this paper to extend the analysis 
of Ref.\ \cite{Martikainen}. We analyze in detail the effects of the geometry for toroidal box potentials, 
as well as those produced 
by the smooth transversal confinement. Interestingly, we show that the vortex dynamics obtained from direct numerical 
simulations can be analytically well described by means of the so--called image method, and conformal mapping techniques.
Finally we analyze the vortex dynamics in presence of dissipation.

% Scheme of the paper

The structure of the paper is as follows. In Sec.\ \ref{sec:model} we describe the physical system, as well as the numerical 
techniques employed. In Sec.\ \ref{sec:box} we consider the case of a toroidal box potential, and introduce the 
image technique, as well as the corresponding conformal mapping, to analytically describe the vortex dynamics, as well as the 
vortex interaction. The case of a toroidal trap with harmonic transversal confinement is considered in Sec.\ \ref{sec:harmonic}, 
where the effects caused by the smooth trapping potential are considered. Sec.\ \ref{sec:dissipation} is devoted to a brief discussion of  
the dissipation effects in the system. We finalize with some conclusions in Sec.\ \ref{sec:conclusions}.

\section{Model}
\label{sec:model}

At sufficiently low temperatures, the dynamics of a condensate in an external potential $V$ can be well described 
by the so--called Gross--Pitaevskii equation \cite{gross61,gross63,pitaevskii61}
\begin{equation}
   i\hbar\frac{\partial\Psi}{\partial t}=
   -\frac{\hbar^2}{2m}\nabla^2\Psi+V\Psi+g|\Psi|^2\Psi,
\label{GP}
\end{equation}
where $g=\frac {4 \pi \hbar^2 a N}{m}$ denotes the coupling constant, $m$ refers to the atomic mass, 
$a$ is the corresponding $s$--wave scattering length, and $N$ is the number of condensed atoms.
We shall consider in the present paper trapping potentials in which the condensate is 
strongly confined along one direction, 
so that the mean field energy is smaller than the excitation energy of the trap in this direction. 
In this situation, the BEC can be considered as quasi--2D, and we can restrict ourselves to the analysis  
of the corresponding 2D GPE, with a modified coupling constant  $g_{2D}=g\int dz |\psi(z)|^4$, where $\psi(z)$ is 
the wavefunction in the strongly confined $z$-direction.
Both 2D \cite{Simo} and quasi--2D \cite{Ketterle,Burger} BEC has been recently observed in experiments.
The potential $V$ in the remaining two dimensions has an annular shape, which we shall consider possessing rigid box boundaries in Sec.\ \ref{sec:box}, or harmonic transversal confinement in Sec.\ \ref{sec:harmonic}. We restrict ourselves in the following to the situations in which the BEC can be considered 
in the Thomas--Fermi regime, i.e. the mean--field energy is much larger than the typical energy spacing of the 
annular trap. We consider in the following numerical simulations the case 
of $N=10^5$ condensed Sodium atoms ($a=2.75$ nm).

We perform the integration of equation (\ref{GP}) 
by an alternating direction implicit (ADI) Crank-Nicholson scheme \cite{press}, using  cylindrical coordinates, 
due to the particular symmetry of the trapping potential. In the present paper, we shall not consider the issue of the 
experimental creation of the vortices, but rather we shall concentrate on the vortex dynamics once created. 
In order to numerically create the vortices, we employ the Crank--Nicholson procedure in 
imaginary time, while ensuring at each time step the proper phase pattern necessary to create the 
vortices \cite{McGee}.  This leads to the creation of vortex states, which are free of 
any other kind of excitation, and, therefore, allow 
for a clear analysis of the vortex dynamics. The proper phase pattern imprinted during 
the imaginary time evolution has 
to be carefully chosen. In particular, the boundary effects have to be taken into account in the method, by 
adding the corresponding phase contribution of the image vortices (see next section).
Once the vortices are created we perform a real time evolution.

\section{Box--like toroidal trap}
\label{sec:box}

In this section, we investigate the evolution of vortex states in an annular trap with rigid boundaries. 
In such a trap the condensate density in the bulk is constant and shrinks to zero at the condensate 
boundaries on a length scale of the order of the healing length $\eta = 1/\sqrt{8 \pi n a}$ ,
where $n$ is the condensate density. Due to the absence of the potential in the interior of the torus 
and the homogeneity 
of the density inside the box, the vortex dynamics is completely determined by the toroidal geometry. In our numerical 
simulations we have considered a trap with inner rigid boundary at a radius $R_1=12\mu$m, 
and an outer one at $R_2=18\mu$m.
We shall not consider, neither in this section nor in the next one, dissipation effects. 
These effects will be discussed 
in Sec.\ \ref{sec:dissipation}.

\subsubsection*{Single vortex dynamics} 

\begin{figure}[htbp]
\begin{center}
\includegraphics[height=50mm,width=75mm,angle=0]{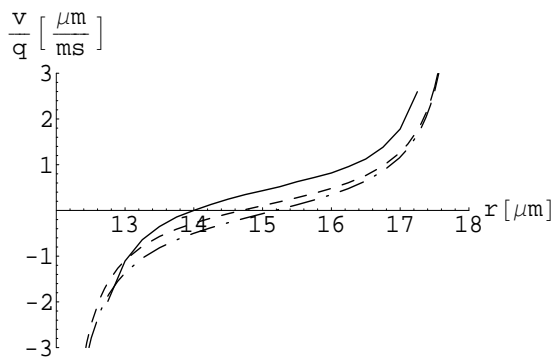}
\end{center}
\captionof{figure}
{Velocity of the vortex in a box like toroidal trap as a function of the radial coordinate.
The three curves are obtained by direct evaluation of Eq. (\ref{velsingle1}) (dashed), Eq. (\ref{velsingle2}) (dashed-dotted) and by solving the GPE numerically  (solid).
}
\label{fig:1}
\end{figure}

We consider first the situation in which a single vortex is placed in a toroidal trapped condensate.
Due to the boundary effects the vortex will not remain at rest, but on the contrary it will move 
at a fixed radius around the torus, where the direction and the modulus of the vortex velocity 
depends on its circulation $\kappa$, and on the radial position at which it is created. 
In Fig.\ \ref{fig:1}, we present the numerical results for the vortex velocity as a function 
of the radial vortex position. 

Such dynamics can be well understood by employing the so--called image method \cite{batchelor67,shlyapnikov99,Guilleumas}. 
Let us briefly review this method.
In a homogeneous (untrapped) condensate the phase of a state containing a 
$q$ times charged vortex is given by $e^{i q \varphi} $ , 
where $\varphi$ is the polar angle centered at the position of the vortex \cite{fetter01}. 
Therefore, the superfluid velocity becomes  
$\vec{v}_{SF}\equiv \hbar\nabla S/m=\hbar q/m\rho\; \vec{e}_{\varphi}$, 
where $S$ is the condensate phase and $\rho$ is the distance from the vortex line.
In the following, we shall only discuss vortices with one quantum of 
circulation since, in the cases considered, vortices with $q>1$ will decay into $q=1$ vortices.
For trapped condensates in the Thomas-Fermi regime, the hydrodynamic form of the Gross-Pitaevskii-equation 
allows to approximately impose for the superfluid velocity field the boundary condition of 
ordinary hydrodynamics, namely the vanishing of the normal component of the superfluid velocity field 
at the condensate boundary \cite{lundh}. 
This boundary condition and the 
specification of the positions and circulations of the vortices, leads to the 
determination of the superfluid velocity field \cite{batchelor67}. 
Moreover, the vortex dynamics in the 
homogeneous, and also in the boxlike trapped quantum gas, reproduces the classical vortex dynamics, 
which is governed by the Kelvin theorem, i.e.  a vortex moves with the same velocity than 
the superfluid at its position \cite{fetter66}.
This velocity can be calculated by employing a well--known method in classical hydrodynamics, namely the replacement 
of the boundaries by additional fictitious vortices, 
so called image vortices. These vortices must be 
placed at the appropriate positions with the appropriate circulations, such that the 
resulting velocity field obeys the boundary 
conditions for the  considered geometry.  

For the toroidal geometry in question, we have to take into account two 
(in principle infinite) families of image vortices. 
For the first family, an image vortex is placed at $\vec{r}_{im1}=\frac{R^{2}_{1}}{r}\frac{\vec{r}}{r}$
with circulation $-\kappa$ and one image vortex with circulation $\kappa$ at the origin \cite{Saffman92}, 
where $\vec{r}$ denotes the position vector of the real vortex in a frame centered at the origin 
of the torus \cite{foot1}. 
The image vortex at $\vec{r}_{im1}$ induces a velocity component
normal on the outer circle, so in addition we need an image vortex at
$\vec{r}_{im2}=\frac{R^{2}_{2}}{r_{im1}} \frac{{\vec{r}}}{r} = \frac{{R^{2}_{2}}}{R^{2}_{1}} \vec{r}$
with circulation $\kappa$.
In turn this image will induce some normal
component on the inner circle, so two more image vortices are required: one at $\vec{r}_{im3}=
\frac{R^{2}_{1}}{r_{im2}} \frac{\vec{r}}{r} = \frac{R^{4}_{1}}{R^{2}_{2}\, r} \frac {\vec{r}}{r}$ with circulation $-\kappa$ the other one at the origin with circulation $ \kappa$  
%{R^{2}_{1}}/{r_{im2}}={R^{4}_{1}}/{R^{2}_{2}}\,r$
and so on and so forth. 
%concerned 
For the second family, one considers an image vortex at 
%$r_{im1}=\frac{R^{2}_{2}}{r}$  
${\vec{r}\,'}_{im1}=\frac {R^{2}_{2}}{r} \frac {\vec{r}}{r}$ with circulation $-\kappa$.
This image vortex gives a normal component of the velocity on the inner
circle so we need in addition one image vortex at
${\vec{r}\,'}_{im2}=\frac {R^{2}_{1}}{r\,'_{im1}} \frac {\vec{r}}{r}= \frac {R^{2}_{1}}{R^{2}_{2}}  {\vec{r}}$
with circulation $\kappa$ as well as one image vortex at the origin with circulation $- \kappa$. For this we require another image vortex at $ {\vec{r}\,'}_{im3}=\frac {R^{2}_{2}}{r\,'_{im2}} \frac {\vec{r}}{r}= \frac {R^{4}_{2}}{R^{2}_{1} \,r} \frac {\vec{r}}{r}$
with circulation $-\kappa$ and so on.

The total superfluid velocity is made up by the contributions of all vortices (both real and image):
\begin{equation}
\vec{v}_{SF}=\sum_{n} q_n \frac{\hbar}{m} \vec{e}_z
\times \frac{\vec{r}-\vec{r}_n}{|\vec{r}-\vec{r}_n|^2}.
\end{equation}
In absence of friction, the superfluid velocity at the vortex position equals the velocity of the vortex line. Therefore we get for the vortex velocity:
\begin{eqnarray}
2\pi \frac{d}{dt}\, \vec{r} & = & \sum ^{\infty }_{n=1}\: \left( \: \frac{\kappa \: \left( r\,- r\, \left( \frac{R_{2}}{R_{1}}\right) ^{2n}\right)\vec{e}_{\varphi}}{\left( \vec{r}-\left( \frac{R_{2}}{R_{1}}\right)^{2n}\vec{r}\right) ^{2}}\,+\, R_{1}\: \leftrightarrow \: R_{2}\;\, \right) \nonumber \\
 & - & \sum ^{\infty }_{n=0}\: \left( \: \frac{\kappa \: \left( r\,\: \left( \frac{R_{1}}{R_{2}}\right) ^{2n}\: \frac{R^{2}_{1}}{r}\:\right) \vec{e}_{\varphi} }{\left( \vec{r}-\left( \frac{R_{1}}{R_{2}}\right) ^{2n}\frac{R^{2}_{1}}{r^{2}}\vec{r}\right) ^{2}}\, +\, R_{1}\: \leftrightarrow \: R_{2}\: \right)\nonumber \\
 & + &  \frac{\kappa}{r}\: \vec{e}_{\varphi}.
\label{velsingle1}
\end{eqnarray}
where $R_{1}\: \leftrightarrow \: R_{2}$ denotes the same term as the precedent one, 
but with $R_1$ and $R_2$ interchanged. 
This velocity field is in good agreement with that calculated from our numerical simulations.

\subsubsection*{Vortex--vortex interaction}

We can at this point analyze the vortex--vortex interaction within an annular trap with 
rigid boundaries. In a homogeneous condensate the vortex--vortex interaction depends on their relative  
circulation. In absence of dissipation, vortices with the same circulation orbit at a fixed distance 
from the central point between them. On the other hand, vortices with opposite circulation move 
parallel to each other forming a so--called vortex pair, which moves as a whole  
with a velocity which depends on the separation between the vortices. 
Two vortices in a toroidal box exhibit a completely different behavior. We shall consider in the following several 
possible scenarios.

\begin{figure}[htbp]
\begin{center}
\includegraphics[height=62mm,width=62mm,angle=0]{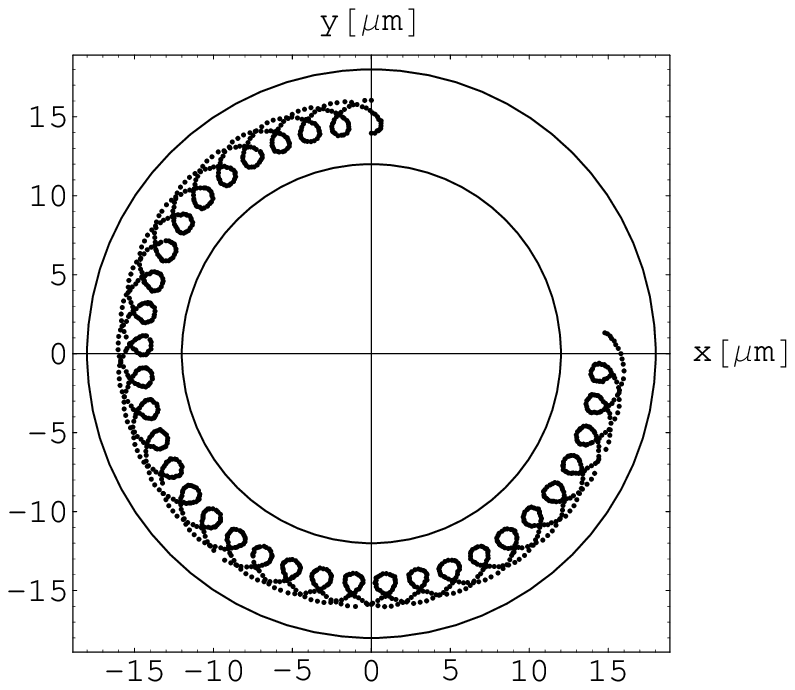}
\includegraphics[height=62mm,width=62mm,angle=0]{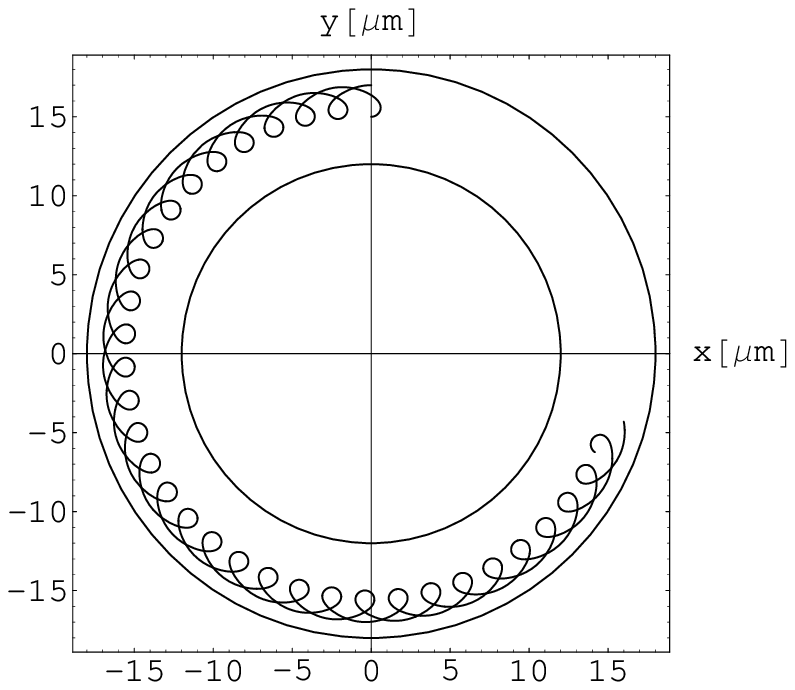}
\end{center}
\captionof{figure}
{(a) Trajectories of vortices with the same circulation created at different radii but on the 
same polar angle, obtained by solving the GPE numerically;  
(b) Results obtained using the image method.}
\label{fig:2}
\end{figure}

\begin{figure}[htbp]
\begin{center}
\includegraphics[height=65mm,width=65mm,angle=0]{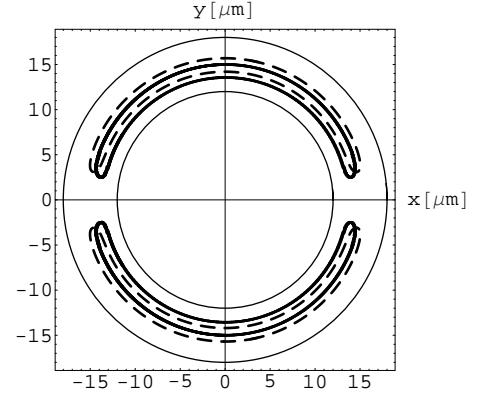}
%\captionof{figure}{Tracectories of a vortex-vortex pair.}
\end{center}
\captionof{figure}
{Trajectories of a vortex-anti vortex pair. The vortices are initially created on the same radial distance but on opposing sides in the torus. We obtain the solid  trajectories by solving the GPE numerically. The dashed trajectories are the computed solutions of the coupled differential equations (\ref{numvelo}).  
The period of oscillation for this setup is 225 ms. } \label{vgloszi}
\label{fig:3}
\end{figure}

\begin{figure}[htbp]
\begin{center}
\includegraphics[height=65mm,width=65mm,angle=0]{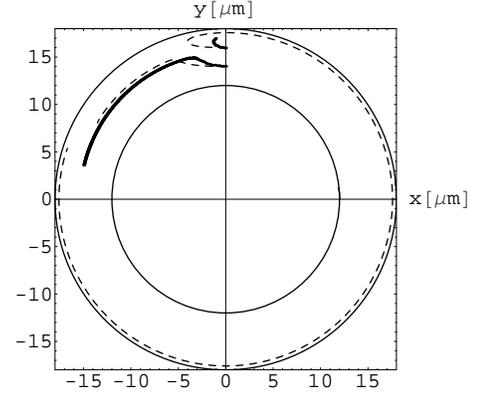}
\end{center}
\captionof{figure}
{Trajectories of a vortex-anti vortex pair. The vortices are initially created on the same polar angle but on different radii. One vortex moves to the condensate boundary, the other one stays in the torus. We obtain the solid trajectories by solving the GPE numerically. The dashed trajectories are the computed solutions of the coupled differential equations (\ref{numvelo}).} 
\label{fig:4}
\end{figure}

Vortices with the same circulation created at different radii but at the 
same polar angle show a dynamics very similar to that of the homogeneous case: they  
perform deformed orbits around each other but with an overall drift around the torus (see. Fig.\ 
\ref{fig:2}). On the contrary, vortices with opposite circulation created on the same radial coordinate, 
but at opposite positions in the torus show an interesting evolution, as already discussed in 
Ref.\ \cite{Martikainen}. 
As shown in Fig.\ \ref{fig:3}, the 
vortices move towards each other keeping the radial coordinate constant. 
When approaching, they move radially, and separate on a new constant radius, 
i.e. effectively they repel each other. 
When they meet at the other side of the torus, they repel each other again, 
and return to the original radial coordinate, 
i.e. the overall trajectory for each vortex is closed. 
If the vortex--anti vortex--pair is created such that the vortices 
are close to each other at the 
same polar angle, one vortex moves very quickly (during a few ms) 
towards the condensate boundary, where it is destroyed. 
The other one moves slightly, changes its initial radial position, and later on moves on a fixed radius 
(Fig.\ \ref{fig:4}). This effect explains the results of  
Ref.\ \cite{Martikainen}, where a dark--soliton created in a toroidal geometry seemed to decay into just one vortex, 
apparently violating the conservation of the angular momentum. The dark--soliton actually decays into 
two vortices, but one of them is rapidly destroyed at the condensate boundary, while the remaining one 
is set into an orbit around the annulus.

Employing, as above, the image method technique, the equation for the position of the $i$--th vortex can be found:
\begin{eqnarray}
& & 2\pi \frac{d}{dt}\, \vec{r}_{i}   =  \nonumber \\
 & & \sum ^{2}_{j=1}\sum ^{\infty }_{n=1}\, \left(  \frac{\kappa _{j} \left( r_{i}\vec{e}_{\varphi_i}- r_{j}\, \left( \frac{R_{2}}{R_{1}}\right)^{2n}\vec{e}_{\varphi_j}\right)} {\left( \vec{r}_{i}-\left( \frac{R_{2}}{R_{1}}\right) ^{2n}\vec{r}_{j}\right) ^{2}} + R_{1}\, \leftrightarrow \, R_{2}\, \right) \nonumber \\
 & - & \sum ^{2}_{j=1}\sum ^{\infty }_{n=0}\, \left(  \frac{\kappa _{j} \left( r_{i}\, \vec{e}_{\varphi _{i}}-\left( \frac{R_{1}}{R_{2}}\right) ^{2n}\: \frac{R^{2}_{1}}{r_{j}}\: \vec{e}_{\varphi _{j}}\right) }{\left( \vec{r}_{i}-\left( \frac{R_{1}}{R_{2}}\right) ^{2n}\frac{R^{2}_{1}}{r^{2}_{j}}\vec{r}_{j}\right) ^{2}} + R_{1} \leftrightarrow  R_{2}\, \right) \nonumber \\
 & + & \; \frac{\kappa _{j\neq i}\: \left( r_{i}\, \vec{e}_{\varphi _{i}\:}-\: r_{j\neq i}\: \vec{e}_{\varphi _{j\neq i}}\right) }{\left( \vec{r}_{i}-\vec{r}_{j\neq i}\right) ^{2}} \quad + \quad \sum ^{2}_{j=1}\frac{\kappa _{j}}{r_{i}}\, \vec{e}_{\varphi _{i}} \nonumber \\\ & & \quad i,j=1,2. \label{numvelo}
\end{eqnarray}
As observed in Figs.\ \ref{fig:2}, \ref{fig:3}, and \ref{fig:4}), the results from Eq. (\ref{numvelo}) 
are in good agreement with the direct numerical simulations. 
Deviations from the numerical solutions of the GPE are basically due to deformations of the vortex core (see 
Sec.\ \ref{sec:conclusions}), since the image method does not take into account the internal 
structure of the vortex core.

\subsection*{Analytical results. Conformal mapping.}

In this section, we obtain the analytical expression of the vortex velocities explicitely, by employing a method which combines the previously presented image method and conformal mapping techniques \cite{batchelor67}.

\subsubsection*{Analytical method}

We introduce a complex velocity potential $p(z=x+iy)=\phi(x,y) + i\gamma(x,y)$, 
such that $dp/dz=u-iw$ is the velocity $\vec{v}=(u,w)$.
$p(z)$ has to be analytic everywhere except for singularities, 
and therefore \(\phi(x,y)    \) and \( \gamma(x,y) \) are 
both solutions to the Laplace equation. One of them can be chosen to
be the real velocity potential : $\nabla\phi=\vec{v}$. 
The curves $\phi =\text{constant}$
and $\gamma =\text{constant}$ intersect in right angles and correspond to the equipotential and stream lines.
It can be easily seen, that
\begin{equation}
 p(z)=-\frac{i\kappa}{2\pi }\log \left( z-z_{j}\right)  
\label{vortpot}
\end{equation}
is the complex velocity potential of a single line vortex at position $z_j$, 
and consequently 
\begin{equation}
\vec{v}=\frac{\kappa }{2\pi }\,\frac{1}{\left| \vec{r}-\vec{r_{j}}\right| 
}\,\vec{e}_{\varphi_j }. 
\end{equation}

In presence of rigid boundaries the velocity field is constrained 
to have a vanishing normal component at the boundaries and, therefore, since the
flow is always in the direction of the stream lines, the
boundaries correspond to curves of constant \( \gamma (x,y) \).
One can always arrange $p(z)$ such that this constant is zero, so one
has $\text{Im} \left[ p(z)\right] =0$ for all $z$ at the boundary, where Im indicates imaginary part.
The potential $p(z)$ is constructed from the Green's function of the corresponding
boundary value problem, by superposing the contribution of the $N$ vortices to the flow \cite{Saffman92}:
\begin{equation}
p(z)=f(z)+\sum ^{N}_{j}\kappa _{j}G(z,z_{j}).
\end{equation}
The term $f(z)$ allows for some further flow due to moving boundaries.
The Green's function has the form $G(z,z_{j})=\text{fund}(z,z_{j})+g(z,z_{j})$,
where  fund$(z,z_{j})$ corresponds to the fundamental solution
without any boundary (and also carries the divergence at the vortex positions). 
According to ($\ref{vortpot}$):  
\begin{equation}
\text{fund}(z,z_{j})=-\frac{i}{2\pi }\log \left( z-z_{j}\right) .
\end{equation}
The function $g(z,z_{j})$ accounts for the constraints and it is smooth 
at the position of the vortices. It can be thought of as the complex
potential induced by the images. 
Therefore, the velocity of the $j$-th vortex is given by:
\begin{eqnarray}
\kappa _{j}\left( u_{j}-iw_{j}\right)&=&\kappa _{j}  \left.\frac{d}{dz}\Big(p(z)-\text{fund}(z,z_j)\Big)\right|_{z=z_{j}} \nonumber \\
&=&\frac{\partial \psi }{\partial y_{j}}+i\frac{\partial \psi }{\partial x_{j}},
\label{velocity}
\end{eqnarray}
where we have employed the, so called, Kirchhoff-Routh function \cite{Saffman92}
\begin{equation}
\psi \equiv 
\sum _{j=1}^{N}
\kappa _{j}\: 
\text{Im}\left [
f(z_{j})+\underset{k\neq j}{\sum ^{N}_{k=1}}\frac{\kappa _{k}}{2}G(z_{j},z_{k})
+\frac{\kappa _{j}}{2}g(z_{j},z_{j})
\right ],
\end{equation}
which is an integral of motion in the case of fixed  boundaries.
Note, that contrary to the situation considered in electrostatics, 
there is no contribution from fund$(z,z_{j})$ at the position
of the vortex, since a vortex (possessing cylindrical symmetry) does not act on itself.

For complicated geometries the velocity potential and the 
Kirchhoff-Routh function are difficult to calculate. 
In order to simplify one can map the complex plane 
which contains the complicated boundary value problem into another complex plane where the 
boundary geometry is simpler.  
A mapping $F(z)$, which carries the $z$-plane into the $\zeta$-plane, 
$\zeta=F(z=x+iy)=s(x,y)+i\; t(x,y)$, 
is called conformal if $F(z)$ is, apart from isolated points, an analytic function of $z$, 
the inverse mapping $z=H(\zeta)$ exists, and $H(\zeta)$ is an analytic function of $\zeta$. 
A conformal mapping maps continuous curves (in particular the boundaries) into continuous curves, 
and the solutions of the Laplace equation into solutions of the Laplace
equation, i.e. if the boundary value problem in the original plane is fulfilled, 
then it is also fulfilled in the new plane with the new boundaries beeing the mappings of the original ones.
The transformation behaviour under the conformal mapping: 
\begin{equation}
\psi =\psi'+\sum _{j}\frac{\kappa ^{2}_{j}}{4\pi }\ln \left|\left.\;\;\; \frac{dH}{d\zeta}\right| _{\zeta=\zeta_{j}}\,\,\right| \label{Kirchhofftranf}
\end{equation}
relates the Kirchhoff-Routh function of the old and new coordinates ~\cite{Saffman92}. This transformation law can be deduced by the relation between the velocity of the j-th vortex in the $z-$plane and in the $\zeta -$plane: 
\begin{eqnarray}
u_j-i\,w_j&=&\frac{d\overline{z}_j}{dt}\nonumber \\&=&\frac{d\overline{\zeta}_j}{dt}
\frac{1}{(dH/d\zeta)|_{\zeta=\zeta_j}}+\frac{i\kappa_j}{4\pi}\frac{\left.
\frac{d^2H}{d\zeta^2}\right|_{\zeta=\zeta_j}}{\left(\left.(dH/d\zeta)\right|_{\zeta=\zeta_j}\right)^2},
\nonumber \\ \label{velotransfo}
\end{eqnarray}
where $z_j$ and $\zeta_j$ denote the position of the j-th vortex in the corresponding coordinates and the bar denotes the complex conjugate.

\subsubsection*{One vortex in the torus}

Let us first calculate the velocity of one vortex in the torus. 
As already shown, the image configuration is slightly complicated. 
Let us now assume that the torus is centered at the origin of the slitted complex z-plane
with radii \( R_{1} \) and \( R_{2} \) respectively. 
The mapping:
\begin{eqnarray}
 F_1(z)=\zeta &=& ( 2 \pi - \theta) +i(\ln (r)-\ln (R_{1})) \nonumber \\
&=&2\pi +i\log z-i\ln R_{1} 
\end{eqnarray}
with \( \theta  \) and r corresponding to the cylindrical coordinates
of the $z$-plane, maps the annulus of the slitted $z$-plane onto a strip of width $b= \ln (R_{2}/R_{1})$ and length 
$2\pi$ in the $ \zeta $ -plane, since we consider here only the principal branch of the logarithm. 
%Taking all branches into account would lead to a complex velocity potential, 
%which has to be expressed in terms of elliptical functions.
The inverse mapping is 
\begin{equation}
z=H_1(\zeta)=R_{1}\exp (-i\zeta ).
\label{invmap}
\end{equation}
We approximate the velocity field in this strip of finite length by the velocity field in a strip of infinite length. 
This approximation obviously becomes better as $R_2/R_1 \rightarrow 1$.
%We calculate the velocity of one vortex in an infinite strip of width $b$ 
%to get the Kirchhoff-Routh function for this configuration. 
Let us consider a second conformal mapping
\begin{equation}
F_2(\zeta)=\xi =\exp \left(\frac{\pi }{b}\zeta\right) , \qquad H_2(\xi)=\zeta=\frac{b}{\pi}\log{\xi}
\label{map2}
\end{equation}
which maps the infinite strip to the upper half plane and the boundaries of the 
infinite strip in the $\zeta$-plane onto the real axis in the \( \xi  \)-plane. 
We denote the position of the vortex in the $\zeta$-plane as $\zeta_{1}$ and in the 
\( \xi  \)-plane as $\xi_1$. In the $\xi$-plane it is sufficient to consider one image vortex 
at position $\overline{\xi}_1$ to fulfill the boundary condition. 
The complex velocity potential in the \( \xi  \)-plane is therefore given by
\begin{equation}
P(\xi)=-\frac{i\kappa}{2\pi }\log(\xi-\xi_1)+\frac{i\kappa}{2\pi }\log(\xi-\overline{\xi}_1).
\end{equation}
The velocity of the vortex in the $\xi$-plane is:
\begin{eqnarray}
U_1^{\xi}-iW_1^{\xi}=\frac{d\overline{\xi}_1}{dt}&=&
\left.\frac{d}{d\xi}\left[P(\xi)+\frac{i\kappa}{2\pi}\log[\xi-\xi_1]\right]\right|_{\xi=\xi_1}\nonumber \\
&=&\frac{i\kappa}{2\pi}\frac{1}{\xi_1-\overline{\xi}_1}.
\end{eqnarray}
According to (\ref{velotransfo}) we calculate the velocity of the vortex in the $\zeta$-plane as:
\begin{equation}
U_1^{\zeta}-i\,W_1^{\zeta}=\frac{\kappa}{4b}\cot[\frac{\pi}{b}t_1].
\end{equation}
From this equation the Kirchhoff-Routh function in the $\zeta$-plane can be obtained by solving (\ref{velocity}):
\begin{equation}
\psi^{\zeta}(t_{1})=\frac{\kappa ^{2}}{4\pi }\ln \left[\sin \left(\frac{\pi }{\ln \frac{R_{2}}{R_{1}}}t_{1}\right)\right],
\end{equation}
where the $\zeta$ denotes that this is the Kirchhoff-Routh function for the $\zeta$-plane.
From the transformation law (\ref{Kirchhofftranf}) we get the Kirchhoff-Routh function of the original
 z-coordinates: 
\begin{equation}
\psi (z_1)=\frac{\kappa ^{2}}{4\pi }\ln \left[\sin \left(\frac{\pi }{\ln \frac{R_{2}}{R_{1}}} 
\ln \left( \frac{r_1}{R_{1}}\right) \right) \right] +\frac{\kappa ^{2}}{4\pi }\ln(r_1).
\end{equation}
And from (\ref{velocity}) we obtain the vortex velocity:
\begin{equation}
\vec{v}=-\,\frac{\kappa}{4r\ln \frac{R_{2}}{R_{1}}} \,\cot \left( 
\frac{\pi }{\ln \frac{R_{2}}{R_{1}}}\ln \left( \frac{r}{R_{1}}\right) \right)\; 
\vec{e}_{\varphi }-\frac{\kappa}{4\pi r}\;\vec{e}_{\varphi }.
\label{velsingle2}
\end{equation}
This analytical result is compared in Fig.\ \ref{fig:1} with the results obtained from the direct summation 
provided by the image method. The agreement is, as observed, excellent.

\subsubsection*{Two vortices in the torus}

Let us now consider two vortices in the torus.
We calculate first the velocity of the $j$-th vortex in an infinite strip of width $b$, 
containing two vortices. After applying again the mapping (\ref{map2}), and following the same notation as 
in the previous section,  the complex potential 
in the new coordinates is made up of the contributions of the two real vortices and two image vortices:
\begin{equation}
P(\xi )=\frac{i\kappa _{1}}{2\pi }\log \left( \frac{\xi-\overline{\xi}_1}{\xi-\xi_1}\right) +\frac{i\kappa _{2}}{2\pi }\log \left( \frac{\xi-\overline{\xi}_2}{\xi-\xi_2 }\right) .
\end{equation}
The velocity of the vortex $1$ in the $\xi$-plane is given by:
\begin{equation}
U_1^{\xi}-iW_1^{\xi} =\frac{i\kappa_1/2\pi}{\xi_1-\overline{\xi}_1}+\frac{i\kappa_2/2\pi}{\xi_1-\overline{\xi}_2}-\frac{i\kappa_2/2\pi}{\xi_1-{\xi}_2}
\end{equation}
According to (\ref{velotransfo}) we calculate the velocity of the vortex in the $\zeta$-plane. After elemetary but tedious calculations we obtain:
\begin{eqnarray}
U_1^{\zeta}-iW_1^{\zeta} & = & \frac{\kappa _{1}}{4b}\cot \left( \frac{\pi }{b}t_{1}\right) \nonumber \\
&& +\frac{\kappa _{2}}{4b}\left[ \frac{\sin \left( \frac{\pi }{b}\left( t_{1}+t_{2}\right) \right) }{\cosh \left( \frac{\pi }{b}\left( s_{1}-s_{2}\right) \right) -\cos \left( \frac{\pi }{b}\left( t_{1}+t_{2}\right) \right) }\right. \nonumber \\
 &  &\quad +\frac{\sin \left( \frac{\pi }{b}\left( t_{2}-t_{1}\right) \right) }{\cosh \left( \frac{\pi }{b}\left( s_{1}-s_{2}\right) \right) -\cos \left( \frac{\pi }{b}\left( t_{2}-t_{1}\right) \right) } \nonumber \\ &&\quad\left. -i\frac{\sinh \left( \frac{\pi }{b}\left( s_{1}-s_{2}\right) \right) }{\cosh \left( \frac{\pi }{b}\left( s_{1}-s_{2}\right) \right) -\cos \left( \frac{\pi }{b}\left( t_{2}-t_{1}\right) \right) }\right.  \nonumber \\& &\quad+\left. i\frac{\sinh \left( \frac{\pi }{b}\left( s_{1}-s_{2}\right) \right) }{\cosh \left( \frac{\pi }{b}\left( s_{1}-s_{2}\right) \right) -\cos \left( \frac{\pi }{b}\left( t_{1}+t_{2}\right) \right) }\right]. \nonumber \\ \nonumber \\
\end{eqnarray}
For the second vortex, we have just to interchange the indices $1$ and $2$.
From (\ref{velocity}) we obtain the Kirchhoff-Routh function:
\begin{eqnarray}
\psi^{\zeta}  & = & \sum ^{2}_{j=1}\frac{\kappa ^{2}_{j}}{4\pi }\ln \left( \sin \left( \frac{\pi }{b}t_{j}\right) \right) \nonumber \\&  &-
\frac{\kappa _{1}\kappa _{2}}{4\pi }\ln \left[ \cos \left( \frac{\pi }{b}\left( t_{1}-t_{2}\right) \right) -\cosh \left( \frac{\pi }{b}\left( s_{1}-s_{2}\right) \right) \right] \nonumber \\
 &  & +\frac{\kappa _{1}\kappa _{2}}{4\pi }\ln \left[ \cos \left( \frac{\pi }{b}\left( t_{1}+t_{2}\right) \right) -\cosh \left( \frac{\pi }{b}\left( s_{1}-s_{2}\right) \right) \right] \nonumber, \\ 
\end{eqnarray}
and from (\ref{Kirchhofftranf}) the Kirchhoff-Routh function for the original z-plane, 
corresponding to two vortices in the torus:
\begin{eqnarray}
\psi  & = & \sum ^{2}_{j=1}\frac{\kappa ^{2}_{j}}{4\pi }\ln \left( \sin \left( \frac{\pi }{b}\ln \frac{r_{j}}{R_{1}}\right) \right)+\sum ^{2}_{j=1}\frac{\kappa ^{2}_{j}}{4\pi }\ln \left( {r_{j}}\right) \nonumber \\&  &-
\frac{\kappa _{1}\kappa _{2}}{4\pi }\ln \left[ \cos \left( \frac{\pi }{b}\ln \left( \frac{r_{1}}{r_{2}}\right) \right) -\cosh \left( \frac{\pi }{b}\left( \theta _{1}-\theta _{2}\right) \right) \right] \nonumber \\
 &  & +\frac{\kappa _{1}\kappa _{2}}{4\pi }\ln \left[ \cos \left( \frac{\pi }{b}\ln\left( \frac{r_{1}r_{2}}{R^{2}_{1}}\right) \right) -\cosh \left( \frac{\pi }{b}\left( \theta _{1}-\theta _{2}\right) \right) \right]. \nonumber \\
\end{eqnarray}
Since the Kirchhoff-Routh function is for fixed boundaries a constant of motion, we obtain from this equation information about the vortex trajectories by computing the hyper surface on which the Kirchhoff function is conserved. In particular we investigate the case of the vortex-antivortex pair, where the vortices are initially placed on the same radial coordinate. Due to symmetry considerations it is clear, that the vortices will always stay on the same radial coordinates $r_1=r_2$  and with this knowledge it is easy to compute the vortex trajectories as curves of constant Kirchhoff-Routh function (see Fig. \ref{kirchtraj}).
\begin{center}
\includegraphics[height=80mm,width=80mm,angle=0]{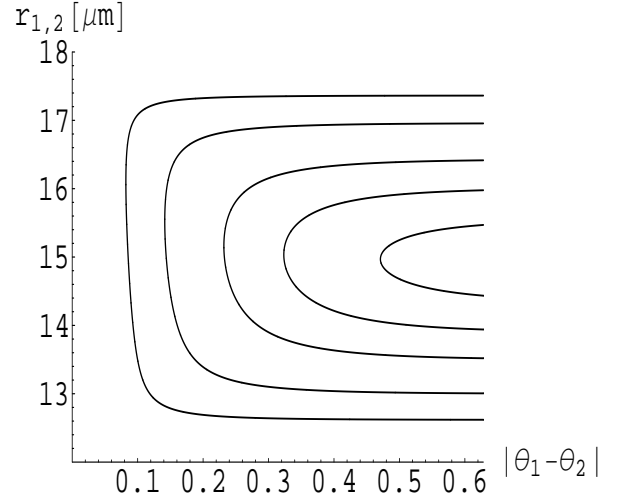}
\end{center}
\captionof{figure}
{Trajectories of a vortex-antivortex pair with vortices placed on the same radial coordinate. On the curves shown the Kirchhoff-Routh function is conserved. The vortices approach, move in radial direction and separate again (see FIG. \ref{vgloszi}). The asymmetry of the image configuration is responsible for the fact, that the trajectories are not symmetric with respect to reflection at $r=15 \mu m$.}\label{kirchtraj} 
\label{fig1}

From the Kirchhoff-Routh function we get the velocity of the $j$-th vortex:
\begin{widetext}
\begin{eqnarray}
\vec{v}_{i} =&  \frac{1}{4\: \ln \left( \frac{R_{2}}{R_{1}}\right) \: r _{i}}\;& \left[                                                       \kappa _{j}\: \frac{\sin \left( \frac{\pi }{\ln \left( \frac{R_{2}}{R_{1}}\right) }\: \ln \left( \frac{r _{i}\: r _{j}}{R^{2}_{1}}\right) \right) }{\cos \left( \frac{\pi }{\ln \left( \frac{R_{2}}{R_{1}}\right) }\: \ln \left( \frac{r _{i}\: r _{j}}{R^{2}_{1}}\right) \right) \: -\: \cosh \left( \frac{\pi }{\ln \left( \frac{R_{2}}{R_{1}}\right) }\: \left( \theta _{i}-\theta _{j}\right) \right) } -\; \kappa _{i}\: \cot \left( \frac{\pi }{\ln \left( \frac{R_{2}}{R_{1}}\right) }\: \ln \left( \frac{r _{i}}{R_{1}}\right) \right) \right.\nonumber \\
 &  &\qquad\left. -\; \kappa _{j}\: \frac{\sin \left( \frac{\pi }{\ln \left( \frac{R_{2}}{R_{1}}\right) }\: \ln \left( \frac{r _{i}}{r _{j}}\right) \right) }{\cos \left( \frac{\pi }{\ln \left( \frac{R_{2}}{R_{1}}\right) }\: \ln \left( \frac{r _{i}}{r_{j}}\right) \right) \: -\: \cosh \left( \frac{\pi }{\ln \left( \frac{R_{2}}{R_{1}}\right) }\: \left( \theta _{i}-\theta _{j}\right) \right) }  -\; \frac{\kappa _{i}}{\pi }\: \ln \left( \frac{R_{2}}{R_{1}}\right)\right] \vec{e}_{\varphi _{i}}
 \nonumber \\
 &  +\; \frac{1}{4\: \ln \left( \frac{R_{2}}{R_{1}}\right) \: r _{i}}&\left[ \kappa _{j}\: \frac{\sinh \left( \frac{\pi }{\ln \left( \frac{R_{2}}{R_{1}}\right) }\left( \theta _{i}-\theta _{j}\right) \right) }{\cos \left( \frac{\pi }{\ln \left( \frac{R_{2}}{R_{1}}\right) }\: \ln \left( \frac{r _{i}}{r _{j}}\right) \right) \: -\: \cosh \left( \frac{\pi }{\ln \left( \frac{R_{2}}{R_{1}}\right) }\: \left( \theta _{i}-\theta _{j}\right) \right) }\right.\nonumber \\
 &  & \left. -\; \kappa _{j}\: \frac{\sinh \left( \frac{\pi }{\ln \left( \frac{R_{2}}{R_{1}}\right) }\left( \theta _{i}-\theta _{j}\right) \right) }{\cos \left( \frac{\pi }{\ln \left( \frac{R_{2}}{R_{1}}\right) }\: \ln \left( \frac{r _{i}\: r_{j}}{R_{1}^{2}}\right) \right) \: -\: \cosh \left( \frac{\pi }{\ln \left( \frac{R_{2}}{R_{1}}\right) }\: \left( \theta _{i}-\theta _{j}\right) \right) }\right] \; \vec{e}_{r _{i}} \nonumber \\
\end{eqnarray}
\end{widetext}
with $i,j = 1,2$, where $i\neq j$. 
In this formula one has to take into account, that the azimutal seperation between the 
vortices fulfills $|\theta_i-\theta_j|\leq \pi$. We have numerically solved the first--order differential 
equations resulting from this velocity field, and obtained the vortex trajectories, whose shapes are practically 
indistinguishable from the ones we obtain by numerically solving (\ref{numvelo}).

\section{Harmonic toroidal trap}
\label{sec:harmonic}

We discuss now the dynamics of vortices in a toroidal geometry with harmonic trapping in the radial direction:
\begin{equation}
V_{trap}(\vec{r})= \frac{1}{2}m\left(\omega_{\rho}^2(\rho-\rho_0)^2+\omega_z^2z^2\right).
\end{equation}
As before we consider a quasi--2D condensate in the Thomas-Fermi-Regime. 
In order to compare with the previous calculation for a box--like annulus, 
the trapping frequencies are adjusted such that the Thomas--Fermi radii correspond to an annular 
condensate with inner radius of $12 \mu m$ and outer radius of $18 \mu m$.
The qualitative dynamic of the vortices is unchanged from that in the box--like case, 
but the time scales are different. Due to the inhomogeneity of the trapping potential 
a vortex will no longer only move with the superfluid velocity. 
In addition it will have some velocity with respect to the superfluid, 
resulting from the small deviation from the cylindrical symmetric vortex state 
caused by the potential. 

\begin{figure}[htbp]
\begin{center}
\includegraphics[height=45mm,width=70mm,angle=0]{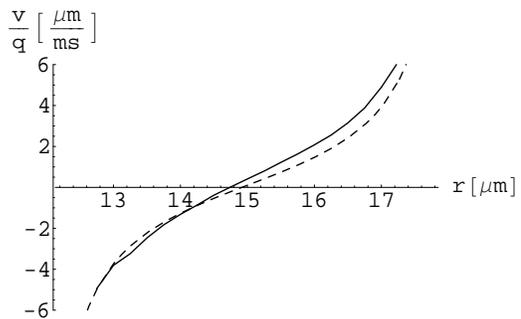}
\end{center}
\captionof{figure}
{Velocity of one vortex in the harmonic toroidal trap as a function of the radial coordinate. 
The solid curve shows the result of the numerical computation of the GPE, the dashed 
curve shows the combination of Eqs. (\ref{velsingle2}) and (\ref{velharm}). }
\label{fig:5}
\end{figure}

This velocity can be calculated by the method of matched asymptotic expansion 
\cite{Rubinstein94,Svidzinsky00}, consisting in evaluating the form of the solution 
of equation (\ref{GP}) in the vortex core and far away from it.
The vortex velocity can be calculated by the requirement that 
the asymptotic behavior of the two solutions has two coincide at a distance of the order of 
the healing length from the center of the vortex. The calculation yields \cite{Svidzinsky00}: 
\begin{eqnarray}
\vec{v}(\vec{r}_0)=-\frac{q\hbar}{2m}\left(\frac{\vec{e}_z\times\nabla V_{trap}(\vec{r}_0)}
{g|\Psi_{TF}|^2}\right)\ln\left(\eta\sqrt{\frac{\nabla^2 V_{trap}}{4g|\Psi_{TF}|^2}}\right), 
\label{velharm} \nonumber \\
\end{eqnarray}
where $\vec{r}_0$ denotes the position of the vortex and $\Psi_{TF}$ is the Thomas--Fermi wave function. 
Since the deviation from the cylindrical symmetric vortex solution in the smooth potential is small,  
we neglect completely their effect on the velocity field induced by the vortex. Therefore, we can employ 
again the image method for determining the velocity field fulfilling the boundary conditions. Therefore, 
the total velocity field results from adding to the field calculated in the previous section that 
of (\ref{velharm}). This produces a very good agreement with our numerical results, 
as observed in Fig.\ \ref{fig:5}.

\begin{figure}[htbp]
\begin{center}
\includegraphics[height=70mm,width=75mm,angle=0]{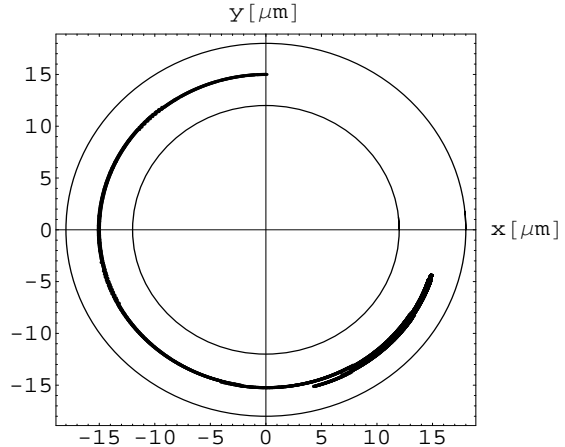}
\end{center}
\captionof{figure}
{Trajectory of one vortex in a box like toroidal trap obtained by numerical solvation of the GPE. We add to the initially flat box like trap a linear, slowly increasing, slope. The vortex slows down as the slope increases and finally turns.}
\label{fig:6}
\end{figure}

The influence of the potential on the vortex velocity gives the possibility to control and manipulate the 
vortex dynamics by adiabatically changing the trapping potential. For instance, it is 
possible to reverse the direction of the vortex orbit in the annular trap 
(Fig.\ \ref{fig:6}). Let us consider the above investigated 
boxlike toroidal trap, where we create one vortex.
%At the beginning of the simulation the depth of the trap is set to $V_0\approx 1\mu K\:k_B$ 
%and one vortex is created. 
If one imposes a slowly increasing linear slope on the initially flat ground of the box
%\begin{equation}
%V_{trap}(r,t)= V_0 [
%\Theta (R_1-r)+\Theta(r-R_2)-\epsilon\Theta(r-R_1)\Theta(R_2-r)(r-R_1)t
%],
%\end{equation}
%where $\Theta$ denotes the step function, and $\epsilon$ is a small parameter, which controls the 
%reversibility of the process. If 
the contribution of (\ref{velharm}) eventually becomes 
more important then the boundary effects, and the vortex turns backwards.

The vortex--vortex interaction is also affected by the transversal harmonic potential, but the qualitative 
behavior described in Sec.\ \ref{sec:box} is maintained \cite{Martikainen}.

\section{Dissipation effects}
\label{sec:dissipation}

\begin{figure}[htbp]
\begin{center}
\includegraphics[height=60mm,width=80mm,angle=0]{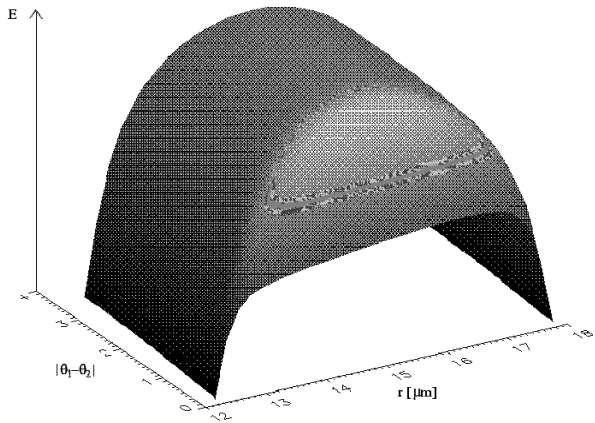}
\end{center}
\caption[]
{Free energy of the condensate containing a vortex-antivortex-pair, 
depending on the positions of the vortices.}
\label{fig:7}
\end{figure}

In this section, we discuss how dissipation affects the vortex dynamics in a toroidal geometry. 
Following a similar analysis as in Ref.\ \cite{Salasnich,McGee} 
we evaluate the free energy functional 
\begin{eqnarray}
E=\int d^3{\bf r} \left \{ \frac{\hbar^2}{2m}|\nabla\psi({\bf r})|^2 +
V({\bf r})|\psi({\bf r})|^2 +\frac{g}{2}|\psi({\bf r})|^4 \right \} \nonumber \\\ \label{energy}
\end{eqnarray}
for a condensate containing 
a vortex-anti-vortex pair with cylindrically 
symmetric vortex cores, for the vortex lines placed at the same radial coordinate $r$, but at different angles $\Theta_1$,$\Theta_2$ (see Fig.\ \ref{fig:7}). 
Since the GPE describes a Hamiltonian evolution and it does not include any dissipation, the trajectories of the vortices according to the GPE  
are curves of constant energy. However, if the vortices approach each other (or the boundaries) to a distance comparable to the healing lenghth, the vortex cores deform and the trajectories we obtain numerically slightly differ from these energy lines.
The total energy increases with increasing distance between the vortices, or between a vortex and the torus 
boundaries, because the kinetic energy grows rapidly in this case.
Therefore, we expect that in presence of dissipation, the vortices will tend to move towards the trap walls 
\cite{shlyapnikov99}. In addition, the vortices will tend to reduce the distance between 
each other, and therefore, if the vortices do not collide with the walls, the apparent repulsion 
introduced by the toroidal geometry will be eventually overcome, and the 
vortices will eventually collide, and mutually annihilate, 
as in homogeneous condensates.

%\begin{figure}[htbp]
%\begin{center}
%\psfig{figure=onevort.eps,width=75mm}
%\includegraphics[height=40mm,width=40mm,angle=0]{120.eps}
%\end{center}
%\caption[]
%{Deformation of the vortex cores when approaching very close to the boundaries of a box--like annular trap. 
%The shaded region denote the region of 95\%--depth of the vortex core.}
%\label{fig:8}
%\end{figure}

\section{Conclusions and outlook}
\label{sec:conclusions}

In this paper, we have investigated the vortex dynamics and vortex--vortex interaction 
in an annular trapping geometry. The vortex trajectories in the presence of rigid boundaries
have been obtained numerically by direct numerical computation of the corresponding 2D GPE. 
We have shown that these trajectories can be 
well described by closed analytical formulas, obtained from the combination of  
the image method and conformal mapping. Additionally, we have considered the case of harmonic 
transversal confinement, and showed that the numerical results are in excellent agreement with the 
analytical ones, once included the effects of the inhomogeneous trapping potential calculated 
from the corresponding matched asymptotic expansion. 

We should note at this point, that the image method is applicable, 
when the vortices are spatially separated by a distance greater than their core size, 
which is of the order of the healing length. The latter statement also concerns 
the distance between the vortices and the boundaries. 
If this is the case, no deformation of the vortex core is produced. 
However, if the vortices approach to distances comparable to the core size, 
the initially cylindrical symmetric vortex cores experience evident deformations and therefore, the results 
provided by the image method begin to depart from the numerically obtained ones. Similarly, 
the dynamics of vortices in the quantum gas begin to differ significantly 
from the classical theory \cite{fetter66,fetter65}.

The analysis of the vortex dynamics has been limited to the case of 2D traps. 
Lower dimensional condensates are currently actively investigated, and very recently 
the experimental observation of 2D (and even 1D) condensates have been reported \cite{Simo,Ketterle,Burger}.
However, the results of the present paper and Ref.\ \cite{Martikainen}, open interesting questions
concerning the behavior of vortices in 3D toroidal condensates.
Preliminar results concerning vortex lines parallel to the torus axis, confirm that in a 3D torus
an essentially similar picture as that discussed for 2D is obtained. However tiltings with respect to the torus axis, 
together with boundary effects, could lead to a significant bending of the vortex lines. This effect will affect the observability of the vortices, since the density minimum will depart from the 
direction of experimental observation. In addition, the bending will significantly distort the vortex 
dynamics, introducing an additional radial velocity to the vortex line. The vortex--vortex interaction 
is also expected to be significantly affected in the presence of a significant bending. 
The analysis of these effects, as well as those related with the propagation 
and interaction of vortex rings in toroidal geometries will be the subject of further investigation.

\acknowledgments

We acknowledge  support from Deutsche Forschungsgemeinschaft (SFB 407),  
TMR ERBXTCT-96-002, and ESF PESC BEC2000+. L. S. wishes to thank the 
Alexander von Humboldt Foundation, the Federal Ministry of Education and 
Research and the ZIP Programme of the German Government. 
Discussions with K.-A. Suominen and J.P. Martikainen are acknowledged.


\begin{thebibliography}{99}
\bibitem{foot1} Since we are dealing with a multiple connected region, the velocity field is not 
determined uniquely by the boundary condition and the specification of the vorticity field alone. 
This is reflected by the fact that the vortex centered at the origin can have arbitrary circulation 
without affecting the boundary condition. We choose the central vortex to have the circulation stated 
in the text, to prevent ring currents. Fixing in that way the circulation of the vortex at the origin, 
sets the so-called cyclic constant of the velocity field 
to zero. With this additional information the velocity field is determined uniquely \cite{batchelor67}.
\end{thebibliography}
\end{document}